# Lumping the Approximate Master Equation for Multistate Processes on Complex Networks

Gerrit Großmann[1], Charalampos Kyriakopoulos[1], Luca Bortolussi[2], and Verena Wolf[1]

[1] Computer Science Department, Saarland University
[2] Department of Mathematics and Geosciences, University of Trieste

**Abstract.** Complex networks play an important role in human society and in nature. Stochastic multistate processes provide a powerful framework to model a variety of emerging phenomena such as the dynamics of an epidemic or the spreading of information on complex networks. In recent years, mean-field type approximations gained widespread attention as a tool to analyze and understand complex network dynamics. They reduce the model's complexity by assuming that all nodes with a similar local structure behave identically. Among these methods the approximate master equation (AME) provides the most accurate description of complex networks' dynamics by considering the whole neighborhood of a node. The size of a typical network though renders the numerical solution of multistate AME infeasible. Here, we propose an efficient approach for the numerical solution of the AME that exploits similarities between the differential equations of structurally similar groups of nodes. We cluster a large number of similar equations together and solve only a single *lumped* equation per cluster. Our method allows the application of the AME to real-world networks, while preserving its accuracy in computing estimates of global network properties, such as the fraction of nodes in a state at a given time.

## 1 Introduction

Various emerging phenomena of social, biological, technical, or economic nature can be modeled as stochastic multistate processes on complex networks [1, 3, 24, 26]. Such networks typically consist of millions or even billions of nodes [1, 3], each one being in one of a finite number of states. The state of a node can potentially change over time as a result of interaction with one of its neighboring nodes. The interactions among neighbors are specified by rules and occur independently at random time points, governed by the exponential distribution. Hence, the underlying process is a discrete-state space Markovian process in continuous time (CTMC). Its state space consists of all labeled graphs representing all possible configurations of the complex network. For instance, in the susceptible-infective (SI) model, which describes the spread of a simple epidemic process, each node can either be susceptible or infected; infected nodes propagate the infection to their susceptible neighbors [19, 5].

Monte-Carlo simulations can be carried out only for small networks [11, 19], as they become very expensive for large networks, due to the large number of simulation runs which are necessary to draw reliable conclusions about the network's dynamics.

An alternative and viable approach is based on mean-field approximations, in which nodes sharing a similar local structure are assumed to behave identically and can be described by a single equation, capturing their mean behavior [18, 3, 4, 10, 12]. The heterogeneous (also called degree-based) mean-field (DBMF) approach proposes a system of ordinary differential equations (ODEs) with one equation approximating the nodes of degree $k$ which are in a certain state [25, 9, 18]. The approximate master equation (AME) provides a far more accurate approximation of the network's dynamics, considering explicitly the complete neighborhood of a node in a certain state [16, 17, 14]. However, the corresponding number of differential equations that have to be solved is of the order $\mathcal{O}\big(k_{\max}^{|\mathcal{S}|}\big)$, where $k_{\max}$ is the network's largest degree and $|\mathcal{S}|$ the number of possible states. A coarser approximation called pair approximation (PA) can be derived from AME by imposing the multinomial assumption for the number of neighbors in a state [16, 17]. Nevertheless, solving PA instead of AME is faster but for many networks not accurate enough [17].

Lumping is a popular model reduction technique for Markov-chains and systems of ODEs [21, 6, 28, 7, 8]. It has also been applied to the underlying model of epidemic contact processes [27, 19] and has recently been shown to be extremely effective for the DBMF equation as well as for the PA approach [20]. In this work, we generalize the approach of [20] providing a lumping scheme for the AME, leveraging the observation that nodes with a large degree having a similar neighborhood structure have also typically very similar behaviors. We show that it is possible to massively reduce the number of equations of the AME while preserving the accuracy of global statistical properties of the network. Our contributions, in particular, are the following: (i) we provide a fully automated aggregation scheme for the multistate AME; (ii) we introduce a heuristic to find a reasonable trade-off between number of equations and accuracy; (iii) we evaluate our method on different models from literature and compare our results with the original AME and Monte-Carlo simulation; (iv) we provide an open-source tool[3] written in Python, which takes as input a model specification, generates and solves the lumped (or original) AME.

The remainder of this paper is organized as follows: In Section 2 we describe multistate Markovian processes in networks and formally introduce the AME. In Section 3 we derive lumped equations for a given clustering scheme and in Section 4 we propose and evaluate a clustering algorithm for grouping similar equations together. Case studies are presented in Section 5. We draw final conclusions and identify open research problems in Section 6.

---

[3] https://github.com/gerritgr/LumPyQest

## 2 The Multistate Approximate Master Equation

In this section, we first define contact processes and introduce our notation and terminology for the multistate AME.

### 2.1 Multistate Markovian Processes

We describe a contact process in a network $(\mathcal{G}, \mathcal{S}, R, L)$ by a finite undirected graph $\mathcal{G} = (V, E)$, a finite set of states $\mathcal{S}$, a set of rules $R$, and an initial state for each agent (node) of the graph $L : V \to \mathcal{S}$. We use $s, s', s''$ and $s_1, s_2, \ldots$ to denote elements of $\mathcal{S}$. At each time point $t \geq 0$, each node $v \in V$ is in a state $s \in \mathcal{S}$. The rules $R$ define how neighboring nodes influence the probability of state transitions. A rule consists of a *consumed* state, a *produced* state, and a transition rate, which depends on the neighborhood of the node. We use integer vectors to model a node's neighborhood. For a given set of states $\mathcal{S}$ and maximal degree $k_{\max}$, the set of all potential neighborhood vectors is $\mathcal{M} = \{\mathbf{m} \in \mathbb{Z}_{\geq 0}^{|\mathcal{S}|} \mid \sum_{s \in \mathcal{S}} \mathbf{m}[s] \leq k_{\max}\}$, where we write $\mathbf{m}[s]$ to refer to the number of neighbors in state $s$.

A rule $r \in R$ is a triplet $r = (s, f, s')$ with $s, s' \in \mathcal{S}, s \neq s'$ and rate function $f : \mathcal{M} \to \mathbb{R}_{\geq 0}$ corresponding to the exponential distribution. A rule $r$ (also denoted as $s \xrightarrow{f} s'$) can be applied at every node in state $s$, and, when applied, it transforms this node into state $s'$. Note that this general formulation of a rule containing the rate function can express all types of rules that are described in [16, 17, 14] such as spontaneous changes of a node's state (independent rules) or changes due to the state of a neighbor (contact rules). The delay until a certain rule is applied is exponentially distributed with rate $f(\mathbf{m})$, with rules competing in a race condition where the one with the shortest delay is executed. This results in an underlying stochastic model described by a CTMC.

In the following, we indicate with $R^{s+} = \{(s', f, s) \in R, s' \in \mathcal{S}\}$ all the rules that change the state of a node into $s$, and with $R^{s-} = \{(s, f, s') \in R, s' \in \mathcal{S}\}$ all rules that change an $s$-node into a different state.

**Example** In the SIS model, a susceptible node can become infected by one of its neighbors. An infected node becomes susceptible again, independently of its neighbors. Hence, the infection rule is $S \xrightarrow{\lambda_1 \cdot \mathbf{m}[I]} I$ and the recovery rule is $I \xrightarrow{\lambda_2} S$, where $\mathbf{m}[I]$ denotes the number of infected neighbors and $\lambda_1, \lambda_2 \in \mathbb{R}_{\geq 0}$ are rule-specific rate constants.

### 2.2 Multistate AME

Here, we briefly present the multistate AME, similarly to [14, 20]. The AME assumes that all nodes in a certain state and with the same neighborhood structure are indistinguishable. We define $\mathcal{M}_k = \{\mathbf{m} \in \mathcal{M} \mid \sum_{s \in \mathcal{S}} \mathbf{m}[s] = k\}$ to be the subset of neighborhood vectors referring to nodes of degree $k$. In addition, for $s_1, s_2 \in \mathcal{S}$ and $\mathbf{m} \in \mathcal{M}$, we use $\mathbf{m}^{\{s_1^+, s_2^-\}}$ to denote a neighborhood vector where all entries are equal to those of $\mathbf{m}$, apart from the $s_1$-th entry, which is equal to $\mathbf{m}[s_1] + 1$, and the $s_2$-th entry, which is equal to $\mathbf{m}[s_2] - 1$.

Let $x_{s,\mathbf{m}}(t)$ be the fraction of network nodes that are in state $s$ and have a neighborhood $\mathbf{m}$ at time $t$, and assume the initial state $x_{s,\mathbf{m}}(0)$ is known. Formally, the AME approximates the time evolution of $x_{s,\mathbf{m}}$ with the following set of deterministic ODEs[4]:

$$\begin{aligned}
\frac{\partial x_{s,\mathbf{m}}}{\partial t} = &\sum_{(s',f,s) \in R^{s+}} f(\mathbf{m}) x_{s',\mathbf{m}} - \sum_{(s,f,s') \in R^{s-}} f(\mathbf{m}) x_{s,\mathbf{m}} \\
&+ \sum_{\substack{(s_1,s_2) \in \mathcal{S}^2 \\ s_1 \neq s_2}} \beta^{ss_1 \to ss_2} x_{s,\mathbf{m}^{\{s_1^+,s_2^-\}}} \mathbf{m}^{\{s_1^+,s_2^-\}}[s_1] \\
&- \sum_{\substack{(s_1,s_2) \in \mathcal{S}^2 \\ s_1 \neq s_2}} \beta^{ss_1 \to ss_2} x_{s,\mathbf{m}} \mathbf{m}[s_1] ,
\end{aligned} \quad (1)$$

where, the term $\beta^{ss_1 \to ss_2}$ is the the average rate at which an $(s, s_1)$-edge changes into an $(s, s_2)$-edge, if $s, s_1, s_2 \in \mathcal{S}$ with $s_1 \neq s_2$.

The first term in the right hand side models the inflow into $(s, \mathbf{m})$ nodes from $(s', \mathbf{m})$ nodes, while the second term models the outflow from $(s, \mathbf{m})$ due to the application of a rule. The other two terms describe indirect effects on a $(s, \mathbf{m})$ node due to changes in its neighboring nodes, again considering inflow and outflow (cf. Fig. 1). In particular, a node in the neighborhood $\mathbf{m}$ of $(s, \mathbf{m})$, say in state $s_1$, changes to state $s_2$ by the firing of a rule.

To compute $\beta^{ss_1 \to ss_2}$ we need to define the subset of rules which consume a $s_1$-node and produce an $s_2$-node: $R^{s_1 \to s_2} = \{(s_1, f, s_2) \in R \mid f : \mathcal{M} \to \mathbb{R}_{\geq 0}\}$. Then

$$\beta^{ss_1 \to ss_2} = \frac{\sum_{\mathbf{m} \in \mathcal{M}} \sum_{(s_1,f,s_2) \in R^{s_1 \to s_2}} f(\mathbf{m}) x_{s_1,\mathbf{m}} \mathbf{m}[s]}{\sum_{\mathbf{m} \in \mathcal{M}} x_{s_1,\mathbf{m}} \mathbf{m}[s]} , \quad (2)$$

where in the denominator we normalize dividing by the fraction of $(s, s_1)$ edges. The total number of equations of AME is determined by the number of states $|\mathcal{S}|$ and the maximal degree $k_{\max}$, and equals:

$$\binom{k_{\max} + |\mathcal{S}|}{|\mathcal{S}| - 1}(k_{\max} + 1) . \quad (3)$$

The binomial arises from the number of ways in which, for a fixed degree $k$, one can distribute $k$ neighbors into $|\mathcal{S}|$ different states, see [20] for the proof.

As $x_{s,\mathbf{m}}$ are fractions of network nodes, the following identity holds for all $t$:

$$\sum_{s,\mathbf{m} \in \mathcal{S} \times \mathcal{M}} x_{s,\mathbf{m}}(t) = 1 \quad (4)$$

---

[4] we omit $t$ for the ease of notation

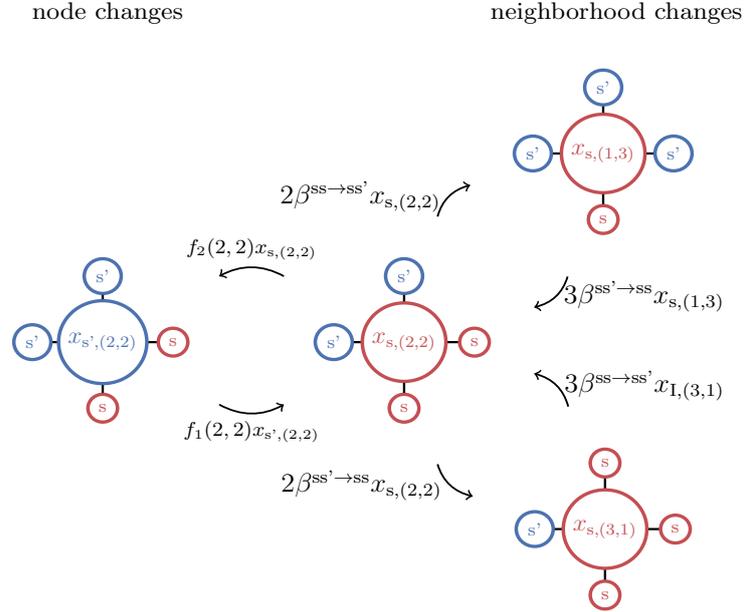

Fig. 1: Illustration of how the AME governs the fraction of $x_{s,(2,2)}$ in a two-state model with rules $(s', f_1, s)$, $(s, f_2, s')$. The inflow and outflow between $x_{s,(2,2)}$ and $x_{s',(2,2)}$ is induced by the direct change of a node's state from $s$ to $s'$ or vice versa. The inflow and outflow between $x_{s,(2,2)}$ and $x_{s,(3,1)}$, $x_{s,(1,3)}$ is attributed to the change of state of a node's neighbor.

Moreover, we use $x_s$ to denote the *global fraction* of nodes in a fixed state $s$, which we get by summing over all possible neighborhood vectors

$$x_s(t) = \sum_{\mathbf{m} \in \mathcal{M}} x_{s,\mathbf{m}}(t) \,, \qquad (5)$$

again with $\sum_{s \in \mathcal{S}} x_s(t) = 1$. Intuitively, $x_s$ is the probability that a randomly chosen node from the network is in state $s$. This is the value of primary interest in many applications, e.g. [3, 24, 26]. Finally, the degree distribution $P(k)$ gives the probability that a randomly chosen node is of degree $k$ ($0 \leq k \leq k_{\max}$). If we sum up all $x_{s,\mathbf{m}}$ which belong to a specific $k$ (i.e. $\mathbf{m} \in \mathcal{M}_k$), as the network structure is assumed to be static, we will necessarily obtain the corresponding degree probability. Hence, for each $t \geq 0$, we have

$$\sum_{s,\mathbf{m} \in \mathcal{S} \times \mathcal{M}_k} x_{s,\mathbf{m}}(t) = P(k) \,. \qquad (6)$$

## 3 Lumping

The key idea of this paper is to group together equations of the AME which have a similar structure and to solve only a single *lumped* equation per group. This lumped equation will capture the evolution of the sum of the AME variables in each group.

Therefore, we divide the set $\{x_{s,\mathbf{m}} \mid s \in \mathcal{S}, \mathbf{m} \in \mathcal{M}\}$ into groups or *clusters*, constructing our clustering such that two equations $x_{s,\mathbf{m}}$, $x_{s',\mathbf{m}'}$ can only end up in the same group if $s = s'$ and $\mathbf{m}$ is 'sufficiently' similar to $\mathbf{m}'$. This ensures that the fractions within a cluster as well as their time derivatives are similar, provided the change in the rate as a function of $\mathbf{m}$ is relatively small when $\mathbf{m}$ is large.

In the sequel, we consider a clustering $\mathcal{C}$ defined as a partition over $\mathcal{M}$, i.e., $\mathcal{C} \subset 2^{\mathcal{M}}$ and $\bigcup_{C \in \mathcal{C}} C = \mathcal{M}$ and all clusters $C$ are disjoint and non-empty. Before we discuss in detail the construction of $\mathcal{C}$ in Section 4, we derive the lumped equations for a given clustering $\mathcal{C}$.

First, recall that we want to approximate the global fractions for each state (cf. Eq. (5)), which can be split into sums over the clusters

$$x_s(t) = \sum_{C \in \mathcal{C}} \sum_{\mathbf{m} \in C} x_{s,\mathbf{m}}(t) \ . \tag{7}$$

Our goal is now to construct a smaller equation system, where the variables $z_{s,C}$ approximate the sum over all $x_{s,\mathbf{m}}$ with $\mathbf{m} \in C$

$$z_{s,C}(t) \approx \sum_{\mathbf{m} \in C} x_{s,\mathbf{m}}(t) \ . \tag{8}$$

Henceforth, we can approximate the global fractions as

$$x_s(t) \approx \sum_{C \in \mathcal{C}} z_{s,C}(t) \ . \tag{9}$$

The number of equations is then given by $|\mathcal{S}| \cdot |\mathcal{C}|$. As one might expect, there is a trade-off between the accuracy of $z_{s,C}(t)$ and the computational cost, proportional to the number of clusters.

### 3.1 Lumping the Initial State and the Time Derivative

As the initial values of $x_{s,\mathbf{m}}$ are given, we define the initial lumped values

$$z_{s,C}(0) = \sum_{\mathbf{m} \in C} x_{s,\mathbf{m}}(0) \ . \tag{10}$$

To achieve the criterion in Eq. (8) for the fractions computed at $t > 0$, we seek for time derivatives which fulfill

$$\frac{\partial z_{s,C}}{\partial t} \approx \sum_{\mathbf{m} \in C} \frac{\partial x_{s,\mathbf{m}}}{\partial t} \ . \tag{11}$$

Note that an exact lumping is in general not possible as $\frac{\partial z_{s,C}}{\partial t}$ is a function of the individual $x_{s,\mathbf{m}}(t)$. In order to close the equations for $z_{s,C}$, we need to express $x_{s,\mathbf{m}}$ as an approximate function of $z_{s,C}$. The naive idea is to assume that the true fractions $x_{s,\mathbf{m}}$ are similar for all $\mathbf{m}$ that belong to the same cluster, i.e., if $\mathbf{m}, \mathbf{m}' \in C$ then $x_{s,\mathbf{m}} \approx x_{s,\mathbf{m}'}$, leading to an approximation of $x_{s,\mathbf{m}}$ as $z_{s,C}/|C|$.

This is however problematic, as it neglects the fact that neighbors of nodes of different degree have different size. In fact, even if for two degrees $k_1 < k_2$ in the same cluster we have $P(k_1) = P(k_2)$ (while typically $P(k_1) > P(k_2)$), the number of possible different neighbors $\mathbf{m}_2$ of a $k_2$-node is larger than the number of different neighbors $\mathbf{m}_1$ of a $k_1$-node, $|\mathcal{M}_{k_1}| < |\mathcal{M}_{k_2}|$, hence typically $x_{s,\mathbf{m}_2} < x_{s,\mathbf{m}_1}$, as the mass of $P(k_2)$ has to be split among more variables. In order to correct for this asymmetry between degrees in each cluster, we introduce the following assumption:

**Assumption**: *All fractions $x_{s,\mathbf{m}}$ inside a cluster $C$ that refer to the same degree contribute equally to the sum $z_{s,C}$. Equations of different degree contribute proportionally to their degree probability $P(k)$ and inversely proportionally to the neighborhood size for that degree.*

Based on the above assumption, we define a degree dependent scaling-factor $w_{C,k} \in \mathbb{R}_{\geq 0}$, which only depends on the corresponding cluster $C$ and degree $k$. According to the above assumption $w_{C,k} \propto \frac{P(k)}{|\mathcal{M}_k|}$. To ensure that the weights of one cluster sum up to one, we define

$$w_{C,k} = \frac{P(k)}{|\mathcal{M}_k|} \cdot \left( \sum_{\mathbf{m} \in C} \frac{P(k_{\mathbf{m}})}{|\mathcal{M}_{k_{\mathbf{m}}}|} \right)^{-1}, \tag{12}$$

where $k_{\mathbf{m}} = \sum_{s \in \mathcal{S}} \mathbf{m}[s]$ is the degree of a neighborhood $\mathbf{m}$. We compute approximations of $x_{s,\mathbf{m}}$ based on $z_{s,C}$ as

$$x_{s,\mathbf{m}} \approx z_{s,C} \cdot w_{C,k_{\mathbf{m}}}. \tag{13}$$

### 3.2 Building the Lumped Equations

To define a differential equation for the lumped fraction $z_{s,C}$, we consider again Eq. (11) and replace $\frac{\partial x_{s,\mathbf{m}}}{\partial t}$ by the l.h.s. of Eq. (1). Then we substitute every occurrence of $x_{s,\mathbf{m}}$ by its corresponding lumped variable multiplied with the scaling factor, i.e., $z_{s,C} \cdot w_{C,k_{\mathbf{m}}}$, where $\mathbf{m} \in C$. Since $\mathbf{m} \in C$ does generally not imply that $\mathbf{m}^{\{s_1^+,s_2^-\}} \in C$, the substitution of $x_{s,\mathbf{m}^{\{s_1^+,s_2^-\}}}$ is somewhat more complicated. Let $C(\mathbf{m})$ denote the cluster $\mathbf{m}$ belongs to. If $\mathbf{m}$ lies "at the border" of a cluster then $C(\mathbf{m}^{\{s_1^+,s_2^-\}})$ might be different than $C(\mathbf{m})$. The lumped AME

takes then the following form:

$$\begin{aligned}
\frac{\partial z_{s,C}}{\partial t} &= \sum_{(s',f,s)\in R^{s+}} z_{s',C}\Big(\sum_{\mathbf{m}\in C} w_{C,k_{\mathbf{m}}}f(\mathbf{m})\Big) \\
&- \sum_{(s,f,s')\in R^{s-}} z_{s,C}\Big(\sum_{\mathbf{m}\in C} w_{C,k_{\mathbf{m}}}f(\mathbf{m})\Big) \\
&+ \sum_{\substack{(s_1,s_2)\in\mathcal{S}^2 \\ s_1\neq s_2}} \beta_{\mathcal{L}}^{ss_1\to ss_2}\Big(\sum_{\mathbf{m}\in C} w_{C(\mathbf{m}^{\{s_1^+,s_2^-\}}),k_{\mathbf{m}}} z_{s,C(\mathbf{m}^{\{s_1^+,s_2^-\}})}\mathbf{m}^{\{s_1^+,s_2^-\}}[s_1]\Big) \\
&- \sum_{\substack{(s_1,s_2)\in\mathcal{S}^2 \\ s_1\neq s_2}} \beta_{\mathcal{L}}^{ss_1\to ss_2} z_{s,C}\Big(\sum_{\mathbf{m}\in C} w_{C,k_{\mathbf{m}}}\mathbf{m}[s_1]\Big) \;,
\end{aligned} \quad (14)$$

where

$$\beta_{\mathcal{L}}^{ss_1\to ss_2} = \frac{\sum_{C\in\mathcal{C}} z_{s_1,C} \sum_{(s_1,f,s_2)\in R^{s_1\to s_2}} \sum_{\mathbf{m}\in C} f(\mathbf{m}) w_{C,k_{\mathbf{m}}}\mathbf{m}[s]}{\sum_{C\in\mathcal{C}} z_{s_1,C} \sum_{\mathbf{m}\in C} w_{C,k_{\mathbf{m}}}\mathbf{m}[s]}. \quad (15)$$

To gain a significant speedup compared to the original equation system, it is necessary that the lumped equations can be efficiently evaluated. In particular, we want the number of terms in the lumped equation system to be proportional to the number of fractions $z_{s,C}$ and not to the number of $x_{s,\mathbf{m}}$. This is possible for Eq. (14), because each time we have a sum over $\mathbf{m}\in C$, for instance $\sum_{\mathbf{m}\in C} w_{C,k_{\mathbf{m}}}f(\mathbf{m})$, we can precompute this value during the generation of the equations and do not have to evaluate it at every step of the ODE solver. The sum

$$\sum_{\mathbf{m}\in C} z_{s,C(\mathbf{m}^{\{s_1^+,s_2^-\}})}\mathbf{m}^{\{s_1^+,s_2^-\}}[s_1]$$

can be evaluated efficiently since we only have to consider lumped variables that correspond to clusters $C(\mathbf{m}^{\{s_1^+,s_2^-\}})$ that are close to $C(\mathbf{m})$, i.e., that can be reached from a state in $C(\mathbf{m})$ by the application of a rule. The number of such neighboring clusters is typically small, due to our definition of clusters, see Section 4.

*Remark 1.* For large $k_{\max}$, the number of neighbor vectors in $\mathcal{M}$, i.e. the size of the AME, becomes prohibitively large. For instance, for a maximum degree of the order of 10 thousands, quite common in real networks, the size of $\mathcal{M}$ becomes of the order of $10^{12}$. Even summing a number of elements of this order while generating equations becomes very costly. To overcome this limit, the solution is to approximate terms involving summations in Eq. (14). Consider for instance $\sum_{\mathbf{m}\in C} w_{C,k_{\mathbf{m}}}f(\mathbf{m})$. Instead of evaluating $f$ at every $\mathbf{m}\in C$ and averaging it w.r.t. $w_{C,k_{\mathbf{m}}}$, we can only evaluate $f$ at the mean neighborhood vector $\langle\mathbf{m}\rangle_C$, where each coordinate is defined as $\langle\mathbf{m}\rangle_C[s] = \sum_{\mathbf{m}\in C} w_{C,k_{\mathbf{m}}}\mathbf{m}[s]$. We can then approximate $\sum_{\mathbf{m}\in C} w_{C,k_{\mathbf{m}}}f(\mathbf{m}) \approx f(\langle\mathbf{m}\rangle_C)$. See Appendix A for details.

## 4 Partitioning of the Neighborhood Set

In this section we describe an algorithm to partition $\mathcal{M}$, and construct the clustering $\mathcal{C}$. Our algorithm builds partitions with a varying granularity to control the trade-off between accuracy and execution speed. We consider three main criteria: the similarity of different equations, their impact on the global error, and how fast is the evaluation of the lumped equations. Furthermore, as the size of $\mathcal{M}$ can be extremely large, we cannot rely on typical hierarchical clustering algorithms having a cubic runtime in the number of elements to be clustered. Our solution is to decouple each $\mathbf{m}$ into two components: its degree $k_\mathbf{m}$ (encoding its length) and its projection to the unit simplex (encoding its direction). We cluster these two components independently.

### 4.1 Hierarchical Clustering for Degrees

Since our clustering is degree-dependent, we first partition the set of degrees $\{0, \ldots, k_{\max}\}$. Let $\mathcal{K} \subset 2^{\{0, \ldots, k_{\max}\}}$ be a degree partitioning, i.e., the disjoint union of all $K \in \mathcal{K}$ is the set of degrees. The goal of the degree clustering is to merge together consecutive degrees with small probability while putting degrees with high probability mainly in separate clusters. This is particularly relevant for the power-law distribution, which is predominantly found in real world networks [2, 1] as it allows us to cluster a large number of high degrees with low total probability all together without losing much information.

We use an iterative procedure inspired by bottom-up hierarchical clustering to determine $\mathcal{K}$. We start by assigning to each degree an individual cluster and iteratively join the two consecutive clusters that increase the cost function $\mathcal{L}$ by the least amount. The cost function $\mathcal{L}$ punishes disparity in the spread of probability mass over clusters, leading to clusters that have approximately the same total probability mass. It is defined as

$$\mathcal{L}(\mathcal{K}) = \sum_{K \in \mathcal{K}} \Big( \sum_{k \in K} P(k) \Big)^2 . \qquad (16)$$

Note that $\mathcal{L}(\mathcal{K})$ is minimal when all $\sum_{k \in K} P(k)$ have equal values. The algorithm needs $\mathcal{O}(k_{\max}^2)$ comparisons to determine the degree cluster of each element. At the end of this procedure, each $\mathbf{m} \in \mathcal{M}$ has a corresponding degree-cluster $K$ with $k_\mathbf{m} \in K$.

### 4.2 Proportionality Clustering

Independently of $\mathcal{K}$, we partition $\mathcal{M}$ along the different components of vectors $\mathbf{m} \in \mathcal{M}$. First, observe that if we normalize $\mathbf{m}$ by dividing each dimension by $k_\mathbf{m}$, we can embed each $\mathcal{M}_k$ into the unit simplex in $\mathbb{R}^{|\mathcal{S}|}$. The idea is then to partition the unit simplex, and apply the same partition to all $\mathcal{M}_k$. More specifically, we construct such partition coordinate-wise. As each element of the normalized $\mathbf{m}$ takes values in $[0, 1]$, we split the unit interval in $p+1$ subintervals $\mathcal{P} = \{[0, \frac{1}{p}), [\frac{1}{p}, \frac{2}{p}), \ldots, [\frac{p-1}{p}, 1]\}$. Then, two normalized neighbor vectors are in the same proportionality cluster if and only if their coordinates all belong to the subinterval $P \in \mathcal{P}$, possibly different for each coordinate.

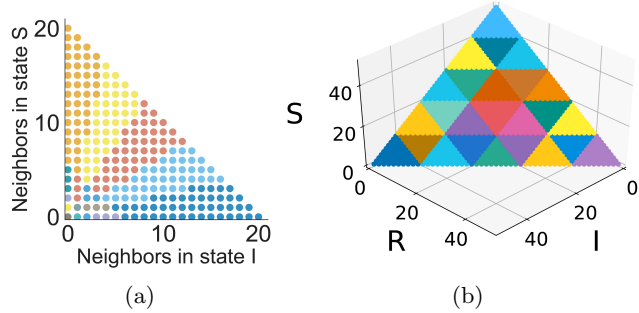

Fig. 2: Left: Clustering of $\mathcal{M}$ for a 2–state model with $k_{\max} = 20$ and $|\mathcal{K}| = |\mathcal{P}| = 7$. Right: Proportionality cluster of a 3–state model with $k_{\max} = 50$ and $|\mathcal{P}| = 5$. Only the plane $\mathcal{M}_{50}$ is shown.

### 4.3 Joint Clusters

Finally, we construct $\mathcal{C}$ such that two points $\mathbf{m}$, $\mathbf{m}'$ are in the same cluster if and only if they are in the same degree-cluster (i.e., $\exists K \in \mathcal{K} : k_{\mathbf{m}}, k_{\mathbf{m}'} \in K$) and in the same proportionality cluster, (i.e., for each dimension $s \in \mathcal{S}$, there exists a $P \in \mathcal{P}$, such that $\frac{\mathbf{m}[s]}{k_{\mathbf{m}}}, \frac{\mathbf{m}'[s]}{k_{\mathbf{m}'}} \in P$).

The effect of combining degree and proportionality clusters, for a model with two different states, is shown in Fig. 2a, where the proportionality clustering gives equally sized triangles that are cut at different degrees by the degree clustering. If we fix a degree $k$, each cluster has only two neighbors (one in each direction). In the 3d-case, the proportionality clustering creates tetrahedra, which correspond to triangles if we fix a degree $k$ (cf. Fig. 2b).

The above clustering admits some advantageous properties: (1) If we fix a degree, all clusters have approximately the same size and spatial shape; (2) The number of 'direct spatial neighboring clusters' of each cluster is always small, which simplifies the identification of clusters in the 'border' cases and eases the generation and evaluation of the lumped AME. Hence, the clusters can be efficiently computed even if $\mathcal{M}$ is very large. Next, we discuss how to choose the size of the clusterings $\mathcal{K}$ and $\mathcal{P}$.

### 4.4 Stopping Heuristic

To find an adequate number of clusters, we solve the lumped AME of the model multiple times while increasing the number of clusters. We stop when the difference between different lumped solutions converges. The underlying assumption is that the approximations become more accurate with an increasing number of clusters and that the respective difference between consecutive lumped solutions becomes evidently smaller when the error starts to level off. Our goal is to stop when the increase in the number of clusters does not bring an appreciable increase in accuracy.

Let $\boldsymbol{z}'(t)$, $\boldsymbol{z}''(t)$ be two solution vectors, i.e., containing the fractions of nodes in each state at time $t$, of the lumped AME that correspond to two different clusterings $\mathcal{C}'$ and $\mathcal{C}''$. We define the difference between two such solutions $\boldsymbol{z}'$, $\boldsymbol{z}''$ as their maximal Euclidean distance over time.

$$\epsilon(\boldsymbol{z}', \boldsymbol{z}'') = \max_{0 \leq t \leq H} \sqrt{\sum_{s \in \mathcal{S}} \left( z'_s(t) - z''_s(t) \right)^2}. \tag{17}$$

For the initial clustering we choose $|\mathcal{K}| = |\mathcal{P}| = c_0$. In each step, we increase the number of clusters by multiplying the previous $c_i$ with a fixed constant, thus $c_{i+1} = \lfloor r c_i \rfloor$ ($r > 1$). We find this to be a more robust approach than increasing $c_i$ by only a fixed amount in each step. We stop when the difference between two consecutive solutions are smaller than $\epsilon_{\text{stop}} > 0$. We consistently observe in all our case studies that $\epsilon(\boldsymbol{z}', \boldsymbol{z}'')$ is a very good indicator on the behavior of the *real* error (cf. Fig. 6b). For our experiments we set empirically $c_0 = 10$, $r = 1.3$, and $\epsilon_{\text{stop}} = 0.01$.

## 5 Case Studies

We demonstrate our approach on three different processes, namely the well-known SIR model, a rumor spreading model, and a SIS model with competing pathogens [22, 13]. We test how the number of clusters, the accuracy, and the runtime of our lumping method relate. In addition, we compare the dynamics of the original and lumped AME with the outcome of Monte-Carlo simulations on a synthetic network of $10^5$ nodes [15, 23]. We performed our experiments on an Ubuntu machine with 8 GB of RAM and quad-core AMD Athlon II X4 620 processor. The code is written in Python 3.5 using SciPy's *vode*[5] ODE solver. The lumping error we provide is the difference between lumped solutions (corresponding to different granularities) and the outcome of the original AME. That is, for the original solution $\mathbf{x}$ and a lumped solution $\mathbf{z}$, we define the lumping errors of $\mathbf{z}$ as $\epsilon(\mathbf{x}, \mathbf{z})$. To generate the error curves, we start with $|\mathcal{P}| = |\mathcal{K}| = 5$ and increase both quantities by one in each step. Note that we test our approach on models with comparably small $k_{\text{max}}$. In general, this undermines the effectiveness of our lumping approach; however using a larger $k_{\text{max}}$ would have hindered the generation of the complete error curve.

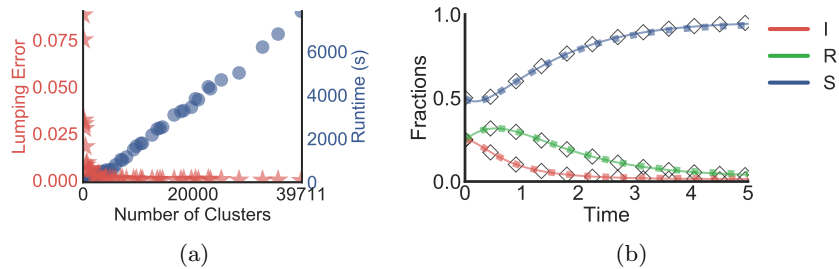

(a)  (b)

Fig. 3: SIR model. (a): Lumping error and runtime of the ODE solver w.r.t. the number of clusters. (b): Fractions of S,I,R nodes over time, as predicted by the original AME (solid line), by the lumped AME (dashed line), and based on Monte-Carlo simulations (diamonds).

### 5.1 SIR

First, we examine the well-known SIR model, where infected nodes (I) go through a recovery state (R) before they become susceptible (S) again:

$$\text{S} \xrightarrow{\lambda_1 \cdot \mathbf{m}[\text{I}]} \text{I} \qquad \text{I} \xrightarrow{\lambda_2} \text{R} \qquad \text{R} \xrightarrow{\lambda_3} \text{S} \ .$$

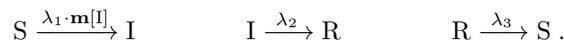

We choose $(\lambda_1, \lambda_2, \lambda_3) = (3.0, 2.0, 1.0)$ and assume a network structure with $k_{\max} = 60$ and a truncated power-law degree distribution with $\gamma = 2.5$. The initial distribution is $(x_\text{I}(0), x_\text{R}(0), x_\text{S}(0)) = (0.25, 0.25, 0.5)$.

In this model the lumping is extremely accurate. In particular, we see that the lumping error of our method becomes quickly very small (Fig. 3a) and that we only need a few hundred ODEs to get a reasonable approximation of the original AME. The lumped solution $\mathbf{z}$ we get from the stopping heuristic, consisting of less than 5% of the original equations, is almost indistinguishable from the original AME solution $\mathbf{x}$ and the Monte-Carlo simulation (Fig. 3b). The lumping error is $\epsilon(\mathbf{x}, \mathbf{z}) = 0.0015$. The lumped solution used here 1791 clusters with a runtime of 235 seconds while solving the original AME we needed 39711 clusters and 7848 seconds.

### 5.2 Rumor Spreading

In the rumor spreading model [13], agents are either ignorants (I) who do not know about the rumor, spreaders (S) who spread the rumor, or stiflers (R) who know about the rumor, but are not interested in spreading it. Ignorants learn about the rumor from spreaders and spreaders lose interest in the rumor when they meet stiflers or other spreaders. Thus, the rules of the model are the following:

$$\text{I} \xrightarrow{\lambda_1 \cdot \mathbf{m}[\text{S}]} \text{S} \qquad \text{S} \xrightarrow{\lambda_2 \cdot \mathbf{m}[\text{R}]} \text{R} \qquad \text{S} \xrightarrow{\lambda_3 \cdot \mathbf{m}[\text{S}]} \text{R} \ .$$

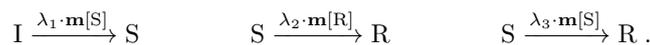

---
[5] https://docs.scipy.org/doc/scipy-0.14.0/reference/generated/scipy.integrate.ode.html

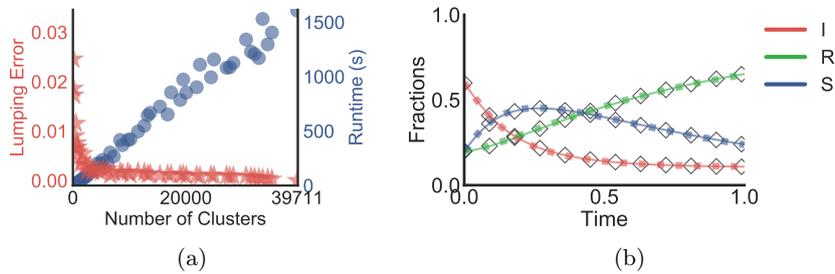

Fig. 4: Rumor spreading model. (a): Lumping error and runtime of the ODE solver w.r.t. the number of clusters. (b): Fractions of nodes in each state over time given by the original AME (solid line), by the lumped AME (dashed line), and based on Monte–Carlo simulations (diamonds).

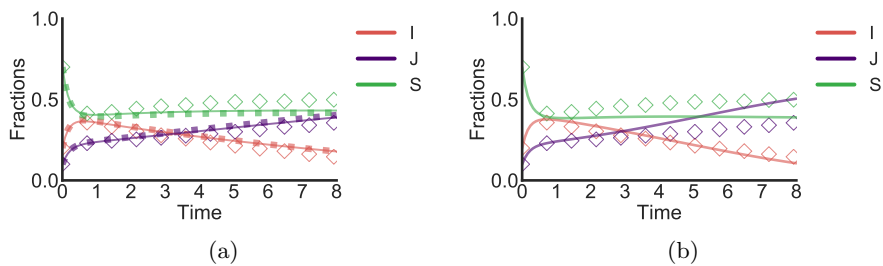

Fig. 5: Competing pathogens dynamics. (a): Fractions of nodes in I, J, S: original AME (solid line); lumped AME (dashed line); Monte-Carlo simulations (diamonds). (b): Comparison of pair approximation with Monte-Carlo simulation (diamonds).

We assume $(\lambda_1, \lambda_2, \lambda_3) = (6.0, 0.5, 0.5)$ with $k_{\max} = 60$ and $\gamma = 3.0$. The initial distribution is set to $(x_I(0), x_R(0), x_S(0)) = (\frac{3}{5}, \frac{1}{5}, \frac{1}{5})$. Again, we find that Monte–Carlo simulations, original AME, and lumped AME are in excellent agreement (Fig. 4b). The error curve, however, converges slower to zero than in the SIR model but it gets fast enough close to it (Fig. 4a). The lumped solution corresponds to 1032 clusters with a lumping error of 0.0059 and a runtime of 35 seconds compared to 39711 clusters of the original AME solution the runtime of which was 1606 seconds.

### 5.3 Competing Pathogens

We, finally, examine an epidemic model with two competing pathogens [22]. The pathogens are denoted by I and J and the susceptible state by S:

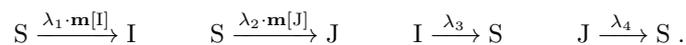

We assume that both pathogens have the same infection rate and differ only in their respective recovery rates. Specifically, we set
$(\lambda_1, \lambda_2, \lambda_3, \lambda_4) = (5.0, 5.0, 1.5, 1.0)$ and assume network parameters of $k_{\max} = 55$ and $\gamma = 2.5$. The initial distribution is $(x_I(0), x_J(0), x_S(0)) = (0.2, 0.1, 0.7)$.

This model is the most challenging case study for our approach. AME solution and naturally lumped AME are not in perfect alignment with Monte Carlo simulations (Fig. 5a) and our lumping approach needs a, comparably to the previous cases, larger number of clusters to get a reasonably good approximation of the AME (Fig. 6a). The computational gain is, however, large as well. The lumped solution that comes with an approximation error of 0.02 corresponds to 2135 clusters and a runtime of 961 seconds compared to 30856 clusters and 17974 seconds of the original AME solution. Pair approximation approach (Fig. 5b) even tough faster (40 seconds) than the lumped AME would have here resulted to a much larger approximation error than our method (cf. Fig. 5b).

At last, the slow convergence of the error curve makes the competing pathogen model a good test case the for our stopping heuristic. The heuristic evaluates the model for three different clusterings (509, 986, 2135 clusters). It stops as the difference between the two last clusterings is smaller than $\epsilon_{\text{stop}}$, showing its effectiveness also for challenging models. In Fig. 6b we show the alignment between the true lumping error and the surrogate error used by the heuristic.

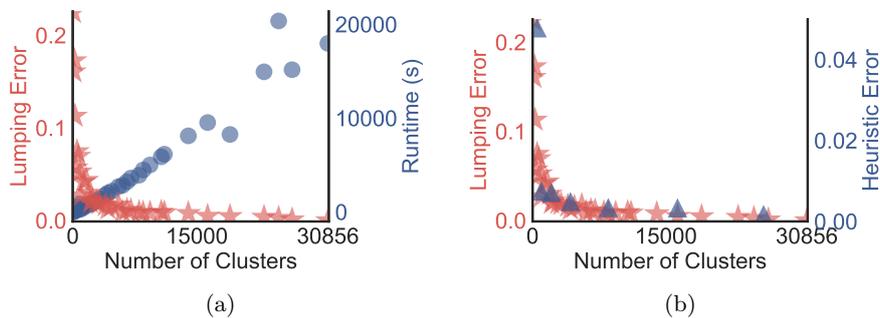

Fig. 6: Competing pathogens lumping. (a): Lumping error and runtime of the ODE solver w.r.t. the number of clusters. (b): Lumping error compared to the error used by the heuristic.

## 6 Conclusions and Future Work

In this paper, we present a novel model-reduction technique to overcome the large computational burden of the multistate AME and make it tractable for real world problems. We show that it is possible to describe complex global behavior of dynamical processes using only an extremely small fraction of the original equations. Our approach exploits the high similarity among the original equations as well as the comparably small impact of equations belonging to the tail of the power-law degree distribution. In addition, we propose an approach for finding a reasonable trade-off between accuracy and runtime of our method. Our approach is particularly useful in situations where several evaluations of the AME are necessary such as for the estimation of parameters or for model selection.

For future work, we plan to develop a method for on-the-fly clustering, which joins equations and breaks them apart during integration. This would allow the clustering to take into account the concrete (local) dynamics and to analyze adaptive networks with a variable degree distribution.

# 7   Acknowledgments

This research was been partially funded by the German Research Council (DFG) as part of the Collaborative Research Center "Methods and Tools for Understanding and Controlling Privacy". We thank James P. Gleeson for his comments regarding the performance of AME on specific models and Michael Backenköhler for his comments on the manuscript.

# References

1. Albert-László Barabási. *Network science*. Cambridge university press, 2016.
2. Albert-László Barabási and Réka Albert. Emergence of scaling in random networks. *science*, 286(5439):509–512, 1999.
3. Alain Barrat, Marc Barthelemy, and Alessandro Vespignani. *Dynamical processes on complex networks*. Cambridge university press, 2008.
4. Luca Bortolussi, Jane Hillston, Diego Latella, and Mieke Massink. Continuous approximation of collective system behaviour: A tutorial. *Performance Evaluation*, 70(5):317–349, 2013.
5. Fred Brauer. Mathematical epidemiology: Past, present, and future. *Infectious Disease Modelling*, 2(2):113–127, 2017.
6. Peter Buchholz. Exact and ordinary lumpability in finite markov chains. *Journal of applied probability*, 31(1):59–75, 1994.
7. Luca Cardelli, Mirco Tribastone, Max Tschaikowski, and Andrea Vandin. Erode: a tool for the evaluation and reduction of ordinary differential equations. In *International Conference on Tools and Algorithms for the Construction and Analysis of Systems*, pages 310–328. Springer, 2017.
8. Luca Cardelli, Mirco Tribastone, Max Tschaikowski, and Andrea Vandin. Syntactic markovian bisimulation for chemical reaction networks. In *Models, Algorithms, Logics and Tools*, 2017.
9. Claudio Castellano and Romualdo Pastor-Satorras. Thresholds for epidemic spreading in networks. *Physical review letters*, 105(21):218701, 2010.
10. Eric Cator and Piet Van Mieghem. Second-order mean-field susceptible-infected-susceptible epidemic threshold. *Physical review E*, 85(5):056111, 2012.
11. Wesley Cota and Silvio C Ferreira. Optimized gillespie algorithms for the simulation of markovian epidemic processes on large and heterogeneous networks. *Computer Physics Communications*, 219:303–312, 2017.

12. G Demirel, F Vazquez, GA Böhme, and T Gross. Moment-closure approximations for discrete adaptive networks. *Physica D: Nonlinear Phenomena*, 267:68–80, 2014.
13. Nick Fedewa, Emily Krause, and Alexandra Sisson. Spread of a rumor. *Society for Industrial and Applied Mathematics. Central Michigan University*, 25, 2013.
14. P. G. Fennell. *Stochastic processes on complex networks: techniques and explorations*. PhD thesis, University of Limerick, 2015.
15. Bailey K Fosdick, Daniel B Larremore, Joel Nishimura, and Johan Ugander. Configuring random graph models with fixed degree sequences. *arXiv preprint arXiv:1608.00607*, 2016.
16. James P Gleeson. High-accuracy approximation of binary-state dynamics on networks. *Physical Review Letters*, 107(6):068701, 2011.
17. James P Gleeson. Binary-state dynamics on complex networks: Pair approximation and beyond. *Physical Review X*, 3(2):021004, 2013.
18. James P Gleeson, Sergey Melnik, Jonathan A Ward, Mason A Porter, and Peter J Mucha. Accuracy of mean-field theory for dynamics on real-world networks. *Physical Review E*, 85(2):026106, 2012.
19. Istvan Z Kiss, Joel C Miller, and Péter L Simon. Mathematics of epidemics on networks: from exact to approximate models. *Forthcoming in Springer TAM series*, 2016.
20. Charalampos Kyriakopoulos, Gerrit Grossmann, Verena Wolf, and Luca Bortolussi. Lumping of degree-based mean-field and pair-approximation equations for multistate contact processes. *Physical Review E*, 97(1):012301, 2018.
21. Genyuan Li and Herschel Rabitz. A general analysis of approximate lumping in chemical kinetics. *Chemical engineering science*, 45(4):977–1002, 1990.
22. Naoki Masuda and Norio Konno. Multi-state epidemic processes on complex networks. *Journal of Theoretical Biology*, 243(1):64–75, 2006.
23. Mark EJ Newman. The structure and function of complex networks. *SIAM review*, 45(2):167–256, 2003.
24. Romualdo Pastor-Satorras, Claudio Castellano, Piet Van Mieghem, and Alessandro Vespignani. Epidemic processes in complex networks. *Reviews of modern physics*, 87(3):925, 2015.
25. Romualdo Pastor-Satorras and Alessandro Vespignani. Epidemic spreading in scale-free networks. *Physical review letters*, 86(14):3200, 2001.
26. Mason Porter and James Gleeson. *Dynamical systems on networks: A tutorial*, volume 4. Springer, 2016.
27. Péter L Simon, Michael Taylor, and Istvan Z Kiss. Exact epidemic models on graphs using graph-automorphism driven lumping. *Journal of mathematical biology*, 62(4):479–508, 2011.
28. James Wei and James CW Kuo. Lumping analysis in monomolecular reaction systems. analysis of the exactly lumpable system. *Industrial & Engineering chemistry fundamentals*, 8(1):114–123, 1969.

## A  Simplification of Equation Generation

We constructed the lumped equations such that they can be evaluated efficiently in each step of the ODE solver. However, for very large $k_{\max}$, the size of $\mathcal{M}$ becomes enormous. This makes the generation of the equations costly because we iterate multiple times over $\mathcal{M}$. This iteration is necessary, each time we compute a scalar of the form $\sum_{\mathbf{m} \in C} w_{C,k_{\mathbf{m}}} f(\mathbf{m})$. In this section we introduce an approximative scheme to generate lumped equations without the computational burden of looking at each individual $\mathbf{m} \in \mathcal{M}$.

Our main idea is to only consider the center of each cluster w.r.t. $w_{C,k_{\mathbf{m}}}$, and not at each cluster element. We use $\langle \mathbf{m} \rangle_C$ to denote the center of cluster $C$, each entry being defined as:

$$\langle \mathbf{m} \rangle_C[s] = \sum_{\mathbf{m} \in C} w_{C,k_{\mathbf{m}}} \mathbf{m}[s] . \tag{18}$$

We can efficiently compute $\langle \mathbf{m} \rangle_C$ by only considering the direction of a cluster (which only depends on the associated proportionality cluster) and the mean degree of a cluster (which can be computed by only considering the degree distribution).

Next, we approximate the average cluster rate by only evaluating the rate function of each rule at the cluster mean:

$$\sum_{\mathbf{m} \in C} w_{C,k_{\mathbf{m}}} f(\mathbf{m}) \approx f(\langle \mathbf{m} \rangle_C) . \tag{19}$$

This, of course, only makes sense if the rate function is reasonably smooth (which is the case in our models).

Likewise, inside the $\beta^{ss_1 \to ss_2}$, we approximate:

$$\sum_{\mathbf{m} \in C} f(\mathbf{m}) w_{C,k_{\mathbf{m}}} \mathbf{m}[s] \approx f(\langle \mathbf{m} \rangle_C) \cdot w_{C,k_{\mathbf{m}}} \cdot \langle \mathbf{m} \rangle_C[s] . \tag{20}$$

Finally, we approximate the in- and outflow related to the $\beta$s. Note that, by design, our clustering has the property that for given $s_1, s_2 \in \mathcal{S}$ and $C \in \mathcal{C}$ there is only exactly one neighboring cluster in which probability mass can flow by adding (resp. subtracting) state $s_1$ (resp. $s_2$) from the neighborhood vector. We now assume a fixed $s_1, s_2$ and use $C' \in \mathcal{C}$ to denote this cluster. We define

$$\begin{aligned} C_{NB} &= \{ \mathbf{m} \mid \mathbf{m}^{\{s_1^+, s_2^-\}} \in C \} \\ C_B &= \{ \mathbf{m} \mid \mathbf{m}^{\{s_1^+, s_2^-\}} \notin C \} = \{ \mathbf{m} \mid \mathbf{m}^{\{s_1^+, s_2^-\}} \in C' \} . \end{aligned} \tag{21}$$

We see that all $\mathbf{m} \in C_{NB}$ occur in the third term (inflow) and in the fourth term (outflow) in the lumped AME. Hence, they cancel out and we can ignore them. To determine the flow of probability mass between two clusters only $C_B$ is of interest. Since the flow is symmetrical (i.e., the inflow of $C'$ is the outflow of $C$) it is sufficient to approximate one direction. We use

$$\beta_{\mathcal{L}}^{ss_1 \to ss_2} \cdot z_{s,C'} \cdot \langle \mathbf{m} \rangle_{C^B}[s_1] \cdot \frac{|C^B|}{|C|} \tag{22}$$

to approximate the flow from $C$ to $C'$. Hence, we add (resp. substract) this value in the equation corresponding to $C'$ (resp. $C$). Note that $\langle \mathbf{m} \rangle_{C^B}$ denotes an approximation of the mean value of $C^B$. As these points lie at the border to $C'$, we use:

$$\langle \mathbf{m} \rangle_{C_B}[s_1] = \frac{1}{2}\langle \mathbf{m} \rangle_C[s_1] + \frac{1}{2}\langle \mathbf{m} \rangle_{C'}[s_1] . \tag{23}$$

$\frac{|C_B|}{|C|}$ is a scaling factor which corresponds to the size of the border between the clusters (the larger the border area, the more probability flows between the clusters). Note that also the cardinality of $C$ and the cardinality of $C_B$ can be efficiently approximated by combinatoric reasoning without looking at the individual elements.

First, consider $|C_B|$. If we fix a $k$, each cluster has approximately the same number of elements (cf. Fig. 2b). We get this number by dividing $|\mathcal{M}_k|$ (the number of neighbor vectors for that degree) by the number of proportionality clusters. To determine $|C|$, we simply aggregate this value over all degrees which occur inside $C$. Next, consider $|C_B|$. It denotes the number of points inside $C$ but next to one particular neighboring cluster. Luckily, our clustering has a nice geometrical structure (namely a triangular one), which we exploit here. The size of a face (surface area in one direction) of each cluster $C \in \mathcal{C}$ in $n$ dimensions for a fixed degree $k$ is exactly the number of elements in a cluster for $n-1$ dimensions for that $k$. For example, the face of a tetrahedron is a triangle, and the size of the triangle can be determined by clustering three dimensions instead of four.

We present two examples of approximative equation generation in Fig. 7. First, we compare these with equations which are generated using the old approach (Fig. 7a). We find that their respective dynamics does not differ significantly.

In addition, we test the approximative equation on a model for which the traditional generation would introduce a significant overhead and where testing the original AME is practically impossible (Fig. 7b). We choose a SIR model (where nodes are trapped in state R) with $k_{\max} = 500$, $\gamma = 2.5$, an infection rate of $3.0 \cdot \mathbf{m}[\mathrm{I}]$, and a recovery rate of 0.3. Again, we see that the lumped AME is in excellent agreement with the numerical simulations.

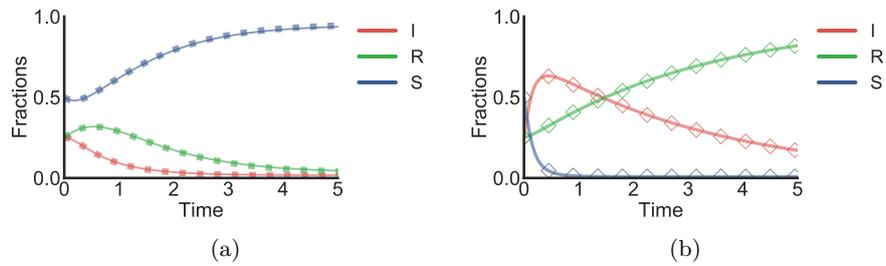

Fig. 7: Dynamics of approximative equations. (a): The same SIR model as in Fig. 3: Lumping of the AME (solid line) and approximation of the lumped equation (dashed line) corresponding to 10 proportionality clusters and 20 degree clusters. (b): New SIR model with 50 degree clusters and 15 proportionality clusters. Approximative equations (solid line) are compared with Monte-Carlo simulations (diamonds). The lumped AME has 8583 clusters. In contrast to more than 21 million clusters of the original model.